\documentclass[a4paper]{spie}  
\usepackage[]{graphicx}

\title{Noise components from black-hole binaries in our galaxy} 

\author{Tomaso M. Belloni\supit{a}
\skiplinehalf
\supit{a}INAF - Osservatorio Astronomico di Brera, Via E. Bianchi 46, I-23807 Merate, Italy
}

\authorinfo{Further author information: 
E-mail: tomaso.belloni@brera.inaf.it, Telephone: 39 039 999 1170}

\begin{document} 
  \maketitle 

\begin{abstract}
Accreting binaries containing a black hole are stellar systems composed of a  normal star and a black hole. Because of the strong gravitational pull of the black  hole, matter is removed from the companion star and falls into the compact object. In falling, it forms an accretion disk of gas that spirals towards the center, heating  up and emitting in X rays. The physics of such a structure is extremely complex and can be studied through observations with X-ray satellites. The time series derived from X-ray observations of bright black-hole binaries in the Galaxy show a complex phenomenology. Broad noise components with a variability of up to $\sim$40\% 
are observed, as well as quasi-periodic features on time scales from 100 seconds down to a few milliseconds. The characteristic frequencies of the different components can change on very short time scales.  However, some of these signals are elusive as they are very weak and are drowned in intrinsic and instrumental noise. The physical nature of these signals is still largely unknown, but it is clear that they originate from gas orbiting a few kilometers from the  central black hole and accreting onto it. In addition of being important for the study of the accretion  of matter onto a black hole, these observational  properties constitute a unique probe for testing General Relativity in the strong  field regime. I review the current observational status as well as the techniques used to study these signals.
\end{abstract}


\keywords{X-ray binaries, black holes, accretion, aperiodic variability, Quasi-Periodic Oscillations}

\section{INTRODUCTION}
\label{sect:intro} 

X-ray astronomy started on June 18 1962, when Geiger counter sensitive to X-rays were mounted on an Aerobee rocket which was launched from New Mexico. The short flight above the Earth's atmosphere, which is opaque to X-rays, led to the detection of the first extra-solar X-ray source, which was named Scorpius X-1. In the next decade, a few more bright sources were discovered, but it was only in 1971 that a satellite for  X-ray astronomy was launched in orbit.  The satellite, called Uhuru, provided the first survey of the sky in X-rays. In addition to many more sources, Uhuru provided the first measurement of intrinsic variability in the flux of a source. The X-rays from Centaurus X-3 were modulated periodically every 4.84 seconds. From Doppler delays and the presence of eclipses in the X-ray flux, it became clear that the pulsar was orbiting in a binary system with another star. This was the first direct evidence of an X-ray binary in the out galaxy: whatever was emitting X-rays was in a binary system with a normal star\cite{tuckergiacconi}.

However, one source seemed to be different from the others. Every new observation, a new different period was found. Although the statistical quality of the data was very poor for modern standards, it was realized that the X-rays from Cygnus X-1 were displaying intrinsic noise, which was tentatively identified with a shot-noise process. Cygnus X-1 was also particularly interesting: the properties of the binary and of its supergiant companion star led to the conclusion that the X-ray emitting object could be a black hole\cite{sewardcharles}.

Since the times of Uhuru, many X-ray satellites were launched, of course with better and better sensitivity and characteristics, as technology advanced. The increase in sensitivity allowed not only to detect weaker and weaker X-ray sources, and in particular to observe X-ray emission from outside our galaxy, but also to obtain better data on the bright ones. In particular, it was realized that, in addition to the bright sources already known, transient sources appear in the X-ray sky. These sources can outshine the persistent ones for time periods of the order of weeks to months, to fade away to very low flux levels, where they can be detected only with the most sensitive instruments. The field of aperiodic timing of astronomical X-ray sources has progressed rapidly together with the availability of more and more sensitive instruments\cite{vdk2006}.

It is important to note that when the signal is stronger than the background, the sensitivity to the detection of broad timing features increases linearly with the source count rate, but only on the square root of the exposure time. Therefore, what is needed is a large instrument that can collect as many photons per unit second as possible, rather than very long exposures. Most of the current knowledge and of the information  in this paper comes from the results of NASA's Rossi X-Ray Timing explorer (RXTE), a satellite for X-ray timing launched at the end of 1995 and still active\cite{swankxte}. Unlike most X-ray missions, which are based on CCD devices and X-ray optics, the main instrument on board RXTE (the Proportional Counter Array, PCA) consists of five large area collimated proportional counters\cite{zhangpca}. For bright sources, they collect a large number of photons, which ensures an unsurpassed performance for timing analysis.

In this article, I will give an overview of our current knowledge of the aperiodic variability of black-hole binaries, i.e. powerful X-ray emitting binary stars where we have evidence that one of the two stars is a black hole. Although we are now detecting similar systems in external galaxies, I will limit myself to the ones on our galaxy, which for distance reasons are the brightest and therefore the ones where we can obtain more precise measurements.
In order to do this, I will outline what we know about the physics of these peculiar (and rare) astronomical objects (Section \ref{sect:xrb}) and present the set of basic analysis tools which are currently used (Section \ref{sect:methods}). The complex observational picture is described in Section \ref{sect:noise}.
Other articles in this session deal with aperiodic variability of binaries containing a neutron star and of powerful active nuclei of external galaxies.

   \begin{figure}
   \begin{center}
   \begin{tabular}{c}
   \includegraphics[height=7cm]{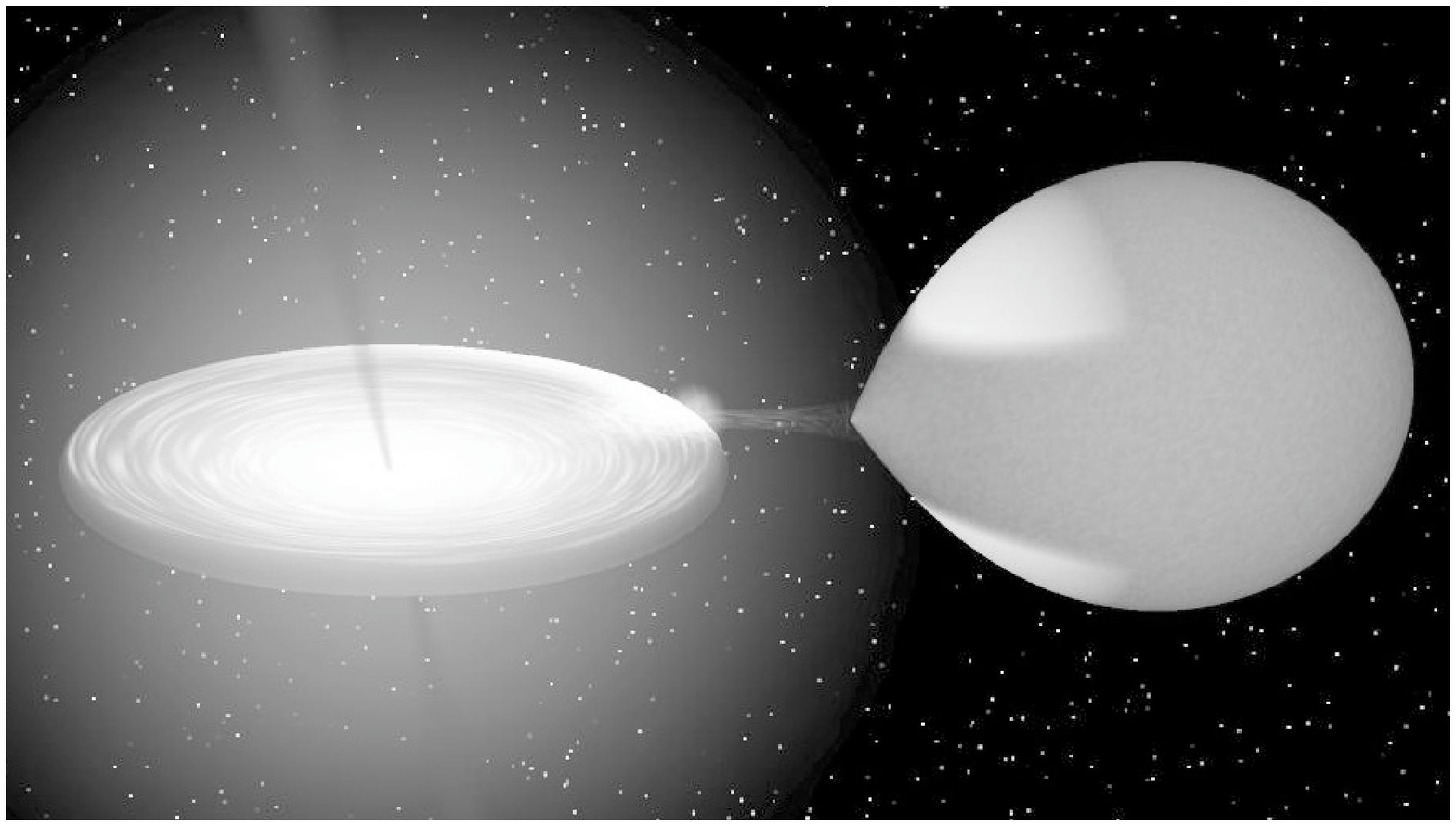}
   \end{tabular}
   \end{center}
   \caption[hynes1655] 
   { \label{hynes} 
Artist impression of an X-ray binary. The mass-losing star is in the right. The compact object (a black hole or a neutron star is at the center of the bright accretion disk. A collimated jet is also visible.
X-ray are emitted from the innermost region of the accretion disk (Credits: R.I. Hynes).}
   \end{figure} 
\section{GALACTIC X-RAY BINARIES}
\label{sect:xrb} 

\subsection{Accretion onto compact objects} 
\label{sect:accretion}

The presence of a compact object, a neutron star of mass 1.4 M$_\odot$ or  a black hole of mass $\sim$10 M$_\odot$ (1 M$_\odot$ is a solar mass, corresponding to $\sim 2\times 10^{33}$ grams), has powerful effects on a binary system. As shown in the drawing of Fig. \ref{hynes}, the strong gravitational pull from the compact object strips matter from the companion. Because of its angular momentum, the gas forms a disk around the compact object, slowly spiraling inwards due to friction effects. The {\it accretion disk} so formed emits radiation whose energy increases inward: the innermost regions, of the order to tens
of kilometers, emit in the X-ray band. As can be seen from Fig. \ref{hynes}, the system is actually more complicated than this: an additional accretion component could be present (a ``corona") and we now know that most of these binaries emit powerful collimated jets that clearly originate from the inner regions of the accretion flow. In addition, in some X-ray binaries, accretion is known to occur through stellar wind capture rather than stripping of gas.

The physics of the accretion process is very complex.  In the case where the compact object is a neutron star, the presence of both a solid surface and a magnetic field of the neutron star strongly affect the properties of the system. In any case, we are dealing with gas orbiting a compact object at a distance of a few dozen kilometers, where the gravitational pull is extreme. This makes these objects unique for the detection of direct measurements of the properties of strong gravitational fields. The theory of General Relativity (GR) has been tested experimentally in the weak field regime (its effects must be taken into account in the modern GPS system for it to be sufficiently precise), but not in the strong field regime. Unfortunately, these are ``dirty'' systems, where a very complex and poorly understood process of accretion is at work. This means that the tools that we have available to study GR effects are not well known, which makes the whole business difficult. Timing analysis constitutes the cleanest approach to this problem, but at present we do not have a successful accepted model for the observational status.

\subsection{Black-hole binaries}
\label{sect:results} 

In this article, I deal with X-ray binaries containing a black hole as compact object. This has important specific characteristics. First, a black hole has no solid surface: we can consider it as a an object with a radius of a few kilometers: matter that crosses that radius disappears in the black hole. This makes a strong difference with a neutron star, where the accreting gas must, if it is not ejected, eventually impact on the surface of the object and release its remaining energy. In addition, while a neutron star can have an ordered magnetic field sufficiently strong to modify the properties of accretion, no such field can be sustained by a black hole, making it a ``cleaner'' system. The effects of solid surface and magnetic field can be seen very clearly in the article by Michiel van der Klis in this volume.
Black holes are the simplest objects in the Universe: they can be completely characterized by three parameters only: their mass, their spin and their charge, which is astrophysically irrelevant. Unfortunately, as already mentioned, the ``accretion tools'' which we have available to study these very clean systems are themselves very dirty.

Most of known black-hole binaries are transient systems. They spend most of the time at very low accretion regimes and display bright outbursts with recurrence times of a few months to centuries. Currently, we know only three systems which are persistently bright. During these outbursts, which 
provide a large number of photons for spectral and timing analysis, their properties are very complex and variable. As far as fast (below a few minutes) variability is concerned, what we observe are broad features in the power spectra. They are traditionally labeled as ``noise'' if their coherence factor is less than 2 and ``Quasi-Periodic Oscillations" (QPO) if it is higher. This is evidently an arbitrary boundary and indeed the time evolution of some components in the power spectra has shown transitions from Q$>$2 to Q$<$2 and vice-versa. 
The most coherent QPOs in black-hole binaries have a coherence of $\sim$10, unlike for neutron-star binaries where values up to $\sim$200 have been observed\cite{vdk2006}.

\section{TIMING ANALYSIS}
\label{sect:methods} 

X-ray detectors count individual photons, recording a combination of arrival time, position and energy, depending on the type of instrument and of telemetry requirements. In addition to source counts, there are background contributions, both instrumental and cosmic. These sources of background, if not highly variable and weaker than the source signal, have the only effect of diluting the source variability and constitute no major problem for the analysis. 
For timing analysis, we are dealing with time series that can be accumulated in different energy bands. In this paper I concentrate on the analysis of single light curves. Therefore, I will not discuss cross-correlation, cross-spectrum and coherence techniques. A more comprehensive description of the techniques outlined here can be found elsewhere\cite{vdkcesme}.

Since we aim at studying broad non-coherent features, which are often variable on time scales shorter than a typical observation, it does not pay to Fourier-transform very long tie series. In addition, for observational reasons related to satellite orbits and/or telemetry and other constraints, typical uninterrupted time series are of the order of 30 minutes to a few hours. Therefore, it is common to produce spectrograms by dividing the time series in segments of length $\sim 10-1000$s, often called ``dynamical power spectra''. Given the low coherence of the expected signal, a boxcar window is used in order to increase statistics. If the features observed in the spectrogram appear to be constant or no features can be detected, the individual spectra are summed to obtain a Welch spectrum, referred to as ``power density spectrum'' (PDS). 

A very important aspect of the analysis is the statistical properties of the estimator, since many signals are rather weak. Therefore, no overlap between data segments is adopted. Moreover, the single PDS are normalized as
$P_j = 2/N_\gamma | a_j |^2$, where $a_j$ are the Fourier coefficients and $N_\gamma$ is the total number of photons in the time series\cite{leahy1983}. This normalization ensures that pure white noise from counting statistics (Poissonian noise) is distributed as a $\chi^2$ with 2 degrees of freedom. 

In the ideal case, the Poisson-counting white noise in the PDS is therefore a simple constant component $P=2$ and could be subtracted away. However, the detectors and the electronics of the satellite introduce a weak correlation between photons, since they have a ``dead time'' after the detection and processing of a photon. This time is usually very small, for the PCA of the order of 10$\mu$s, but at lower frequencies it has the effect if lowering the Poissonian-noise level of a roughly constant amount depending on the source intensity, but its precise functional dependence is rather complicated. The deadtime of the PCA is well studied, but it still constitutes a limitation for the detection of weak features. 

In the past years, it has become common to represent the PDS with a different normalization. After the Poissonian noise contribution has been subtracted, the power is converted to squared fractional rms, basically dividing by the square of the source count rate\cite{bhatlas}. This makes different PDS corresponding to different fluxes or sources comparable.
In order to average power over a number of bins, the PDS is usually rebinned logarithmically in frequency (each frequency bin is a factor of 1+$\epsilon$ wider than the previous one). The typical representation of the PDS in log-log space.
Finally, recently it has become common practice to plot the PDS after having multiplied power (or squared fractional rms) by frequency\cite{belloni97}. This allows to have comparable power per decade in frequency and therefore estimate the relative amount of variability of the different components. Moreover, in case of broad Lorentzian components, this so-called $\nu P_\nu$ representation has the advantage of peaking at $\nu_M = \nu_0 \sqrt{1+1/(4Q^2)}$, where $\nu_0$ is the centroid frequency and $Q$ is the coherence factor. For high $Q$ Lorentzians, $\nu_M = \nu_0$ and for broad components it tends to the half width at half maximum of the Lorentzian\cite{bpk}.

\section{APERIODIC VARIABILITY: NOISE AND QPO}
\label{sect:noise} 

The properties of fast time variability of black-hole binaries are very complex and are associated to
different energy distributions of the X-ray emission. Since the first studies, these properties have been
classified into separate ``states", which have seen different definitions about which there is
still no consensus. What is important here is that different states are identified mostly, but not exclusively, by the properties of fast variability.
Here, I make use of one of the state definition proposed in the recent years\cite{hombel05,bel3392005,belcefalu}.
It is important to note that the dynamical time scales in the innermost regions of the accretion flow
around a black hole, those where high-energy emission originates, are of the order of 1-10 ms, while
the frequencies described in this sections are considerably lower (100 ms to minutes). Therefore, 
the physical origin of this variability cannot be attributed to simple dynamical effects. 
In the following, I outline briefly the basic observational properties in the different states, concentrating on the aperiodic timing. 

\subsection{Low-Hard State (LS): very strong noise} 
\label{sect:ls}

\begin{figure}
   \begin{center}
   \begin{tabular}{c}
   \includegraphics[height=7cm]{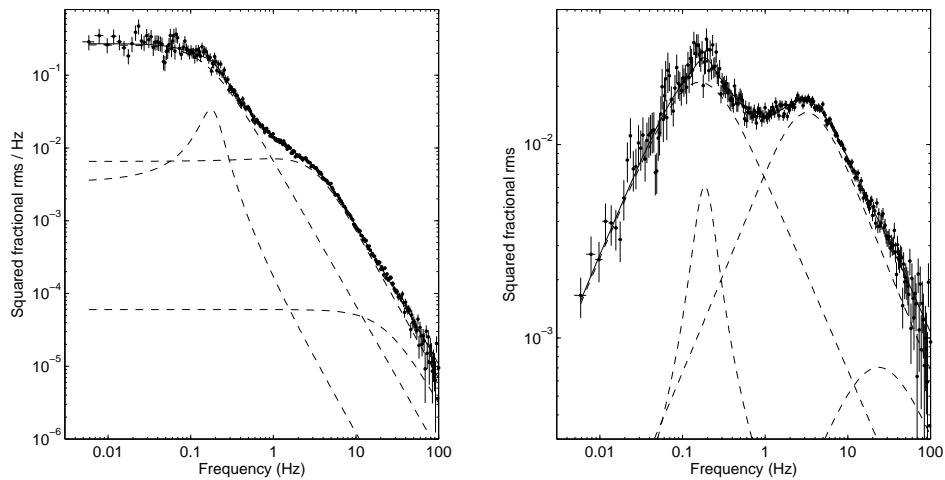}
   \end{tabular}
   \end{center}
   \caption[ls_pds] 
   { \label{ls_pds} 
Power Density Spectrum of GX 339-4 from a RXTE observation of 2002 April 21. The dashed lines
indicate the different Lorentzian components used to fit the overall PDS. Left panel: conventional
representation in squared fractional rms per unit frequency. Right panel: the same PDS, but
multiplied by frequency.
}
   \end{figure}

This state is characterized by a hard energy spectrum, usually interpreted
as the result of repeated inverse Compton scatterings of soft ($< 1$keV) photons by a population of high-energy ($\sim 100$ keV) thermal electrons. The soft photon input is supposed to come from the 
accretion disk, which is observed to show little variability (see below). The strong variability seen in the LS must therefore originate from the scattering gas itself. Figure \ref{ls_pds} shows the PDS of a typical observation of a source in the LS, the bright transient GX 339--4. The shape of this PDS is a textbook example for the LS. It can be decomposed into four components: three flat-top zero-centered Lorentzians plus an additional peaked Lorentzian for a weak and broad QPO. The integrated fractional rms is around 40\%.
It was only a few years ago that it was realized that the PDS of sources in the LS can be decomposed in a number of Lorentzian components, most of which with zero centroid frequency\cite{nowaklorentzians,bpk}. Since this is the same model usually adopted for QPO peaks, it is possible to compare the properties of all components. Unfortunately, as will be shown below, not all states show PDS that can be fitted with such a model.

Although this is the typical shape, the characteristic frequencies (see Sect. \ref{sect:methods}) of the different components vary. In transient systems, where the luminosity variations in this state are large, the frequencies of all these components increase with source luminosity. For the only persistent system know, Cygnus X-1, one particular component seems to stay at the same frequency
\cite{pottschmidttiming}.
In particular, since 1990 is has been noticed that the variation of the flat-top level of the lowest-frequency Lorentzian and its characteristic frequency are anti-correlated in a way to preserve the total fractional rms\cite{bhcygx1}. The total fractional rms decreases with increasing frequencies (and luminosity) due to variations of the other components. The fractional rms decreases slightly if photons of higher energy are selected.

   \begin{figure}
   \begin{center}
   \begin{tabular}{cc}
   \includegraphics[height=7cm]{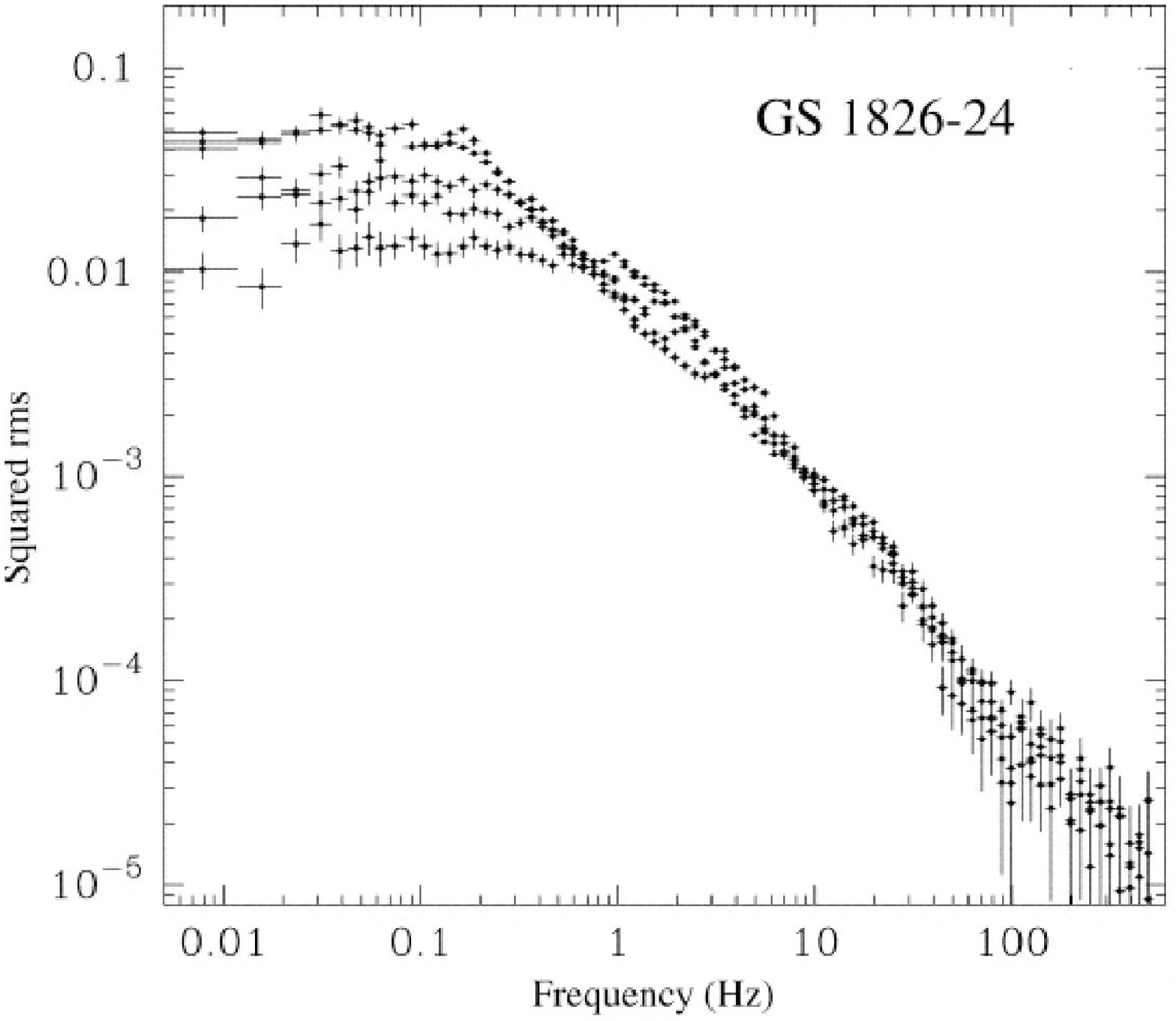} & \includegraphics[height=7cm]{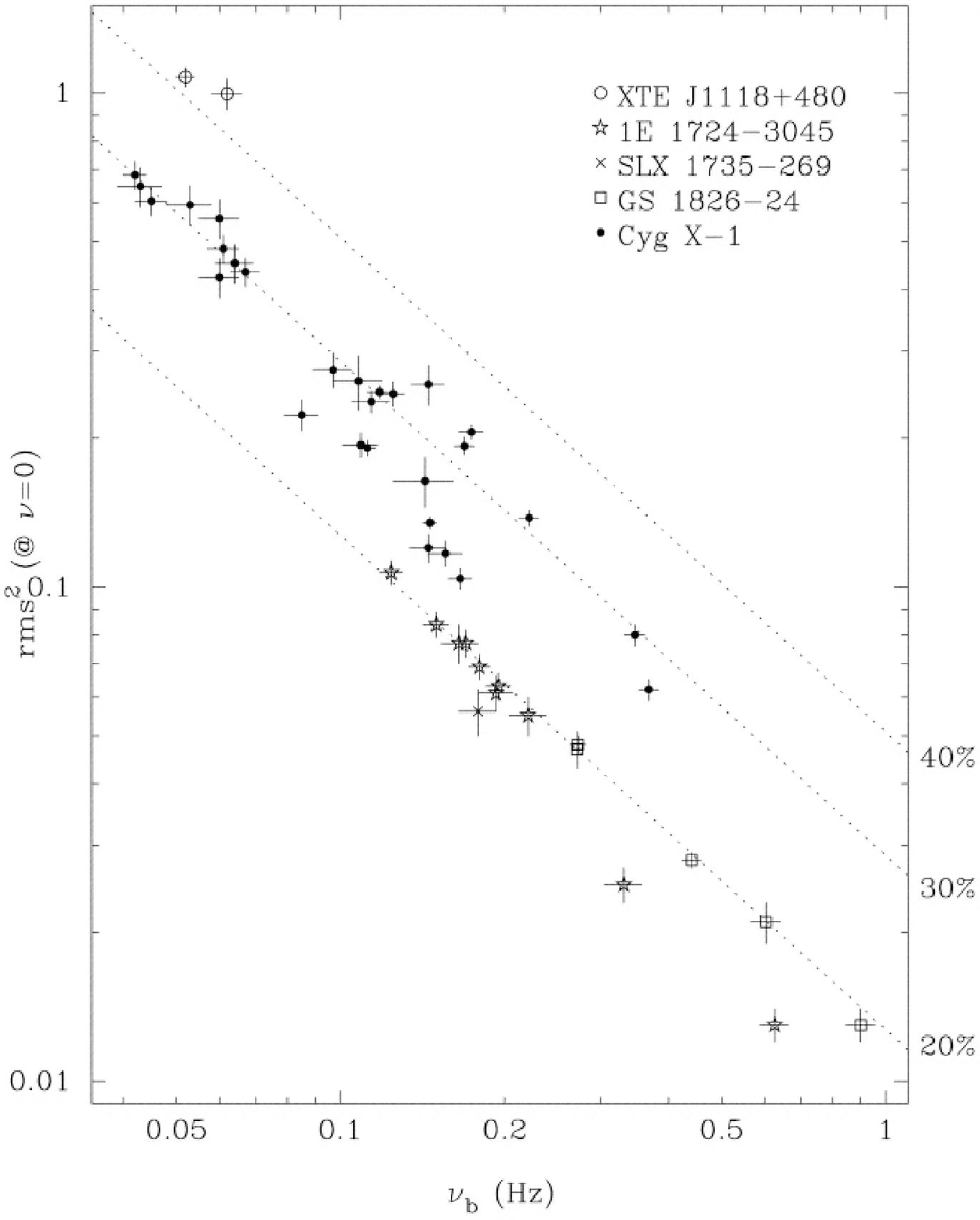}\\
   \end{tabular}
   \end{center}
   \caption[bh_effect] 
   { \label{bh_effect} 
Left panel: a set of PDS from different RXTE observations of the transientr GS 1826--24. Right
panel: correlation between flat-top level and characteristic frequency of the lowest-frequency Lorentzian component in the PDS of sources in low/hard state. The diagonal lines indicate constant integrated fractional rms. Both figures are from Ref. \citenum{bpk}.
}
   \end{figure} 
 
 \subsection{High-Soft State (HS): very weak noise} 
\label{sect:hs}

\begin{figure}
   \begin{center}
   \begin{tabular}{cc}
   \includegraphics[height=6.5cm]{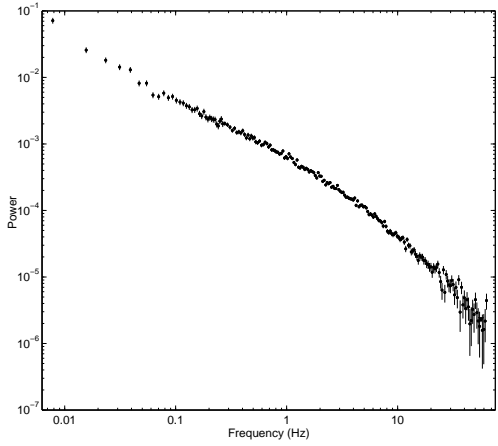} & \includegraphics[height=6cm]{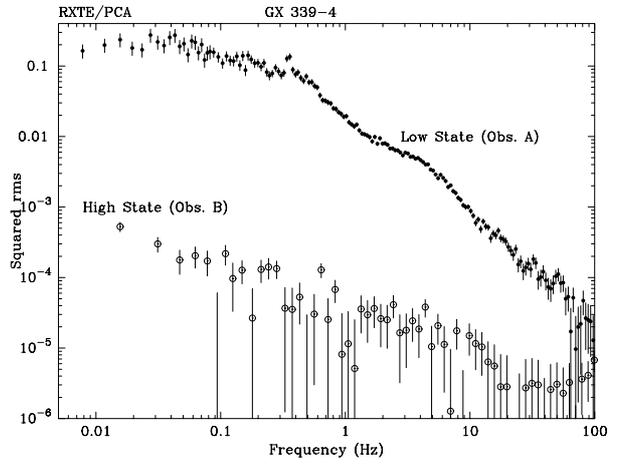}\\
   \end{tabular}
   \end{center}
   \caption[hs_pds] 
   { \label{hs_pds} 
Left panel:  PDS from the bright transient GRO J1655--40 from 2005 March 19, when the source was in the high-soft state. The integrated 0.01--64 Hz fractional rms is $\sim$5\%.
Right panel: comparison PDS from the LS (fractional rms 41\%) and HS (fractional rms 2\%) for GX 339-4 in 1998\cite{belloni3391999}.
}
   \end{figure}

The energy spectrum in this state is completely different from that of the LS. The flux is dominated by a thermal component attributed to the accretion disk around the black hole. As for the LS, the HS can be observed over a wide range of source luminosities, but unlike the LS it is never observed at the beginning and at the end of an outburst. For this state there is a theoretical model that can successfully explain the emitted spectrum\cite{shakurasunyaev}: very little variability is expected and indeed very little variability is observed, of the order of 1--5\%. There is an additional weak spectral component contributing to the flux, which might be responsible for the entire observed variability, but it is difficult to estimate its contribution due to its weakness. When it is possible to measure it, the PDS is well approximated by a power law with an index of $\sim$1-1.5 and a high-frequency cutoff. Sometimes, a QPO peak at around 10--20 Hz is seen. Clearly, this is the least interesting state for time series analysis. What is needed is high sensitivity at high energies, above 10 keV, which is very difficult to achieve instrumentally.
 An example of PDS from the source GRO J1655--40 in its 2005 outburst can be seen in the left panel of Fig. \ref{hs_pds}. The quality of the PDS is good, due to the relatively high ($\sim$5\%) fractional rms, the source brightness and the long exposure time. However, this PDS corresponds to photons in the 3--15 keV range. At higher energies the signal is too weak. 
A comparison between the PDS of the LS and the HS of the same source (GX 339--4)
\cite{belloni3391999} can be seen in the right panel of Fig. \ref{hs_pds}. The fractional rms for the two PDS is 41\% and 2\%.
 
\subsection{Hard intermediate state (HIMS): QPOs with noise} 
\label{sect:hims}

\begin{figure}
   \begin{center}
   \begin{tabular}{c}
   \includegraphics[height=7cm]{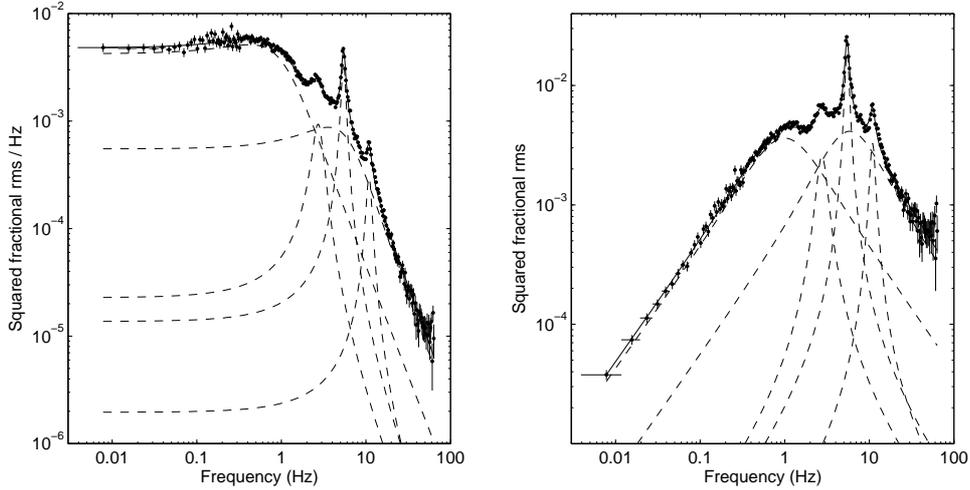}\\
   \end{tabular}
   \end{center}
   \caption[hims_pds] 
   { \label{hims_pds} 
Left panel:  PDS from the bright transient GX 339--4 from 2002 May 11, when the source was in the Hard-Intermediate state. The integrated 0.01--64 Hz fractional rms is $\sim$15\%.
The dashed lines indicate the different Lorentzian components used to fit the overall PDS. Right panel: the same PDS, but multiplied by frequency.
}
   \end{figure}

It is now clear that the transitions between the hard and soft state can be considered separate states and are the most interesting in terms both of high-energy properties and of their relation to the ejection of relativistic jets\cite{fbg2004}. The Hard Intermediate state (HIMS) can be considered an extension of the Low-Hard state. The observed flux is a combination of the components dominating in the LS and the HS, hence the definition ``intermediate'', with varying relative contribution. The PDS is a version of the one in the LS, shifted to higher frequencies and to lower fractional rms. An example can be seen in Fig. \ref{hims_pds}, which shows the PDS of GX 339--4 only 20 days after that of Fig. \ref{ls_pds}. Here a good fit is obtained with two broad noise components plus three QPO peaks. The QPOs are harmonically related to each other: the frequencies are in the ratio 1:2:4. Sometimes a fourth peak appears at a frequency 8 times the first. The harmonic structure of this signal is octave-based. Since the second peak is usually the most intense, it is that one which is considered the fundamental, while the lower one is called a 'sub-harmonic' (see e.g. \citenum{casellaabc}. With this identification, the different Lorentzians can be identified with the same components of the LS. In other words, the components are the same, but at higher frequencies and with shapes that can be different.
As for the LS components, the frequencies of all Lorentzians in the HIMS move together in frequency: while in the LS the frequencies increased with luminosity, here they increase as the source energy spectrum softens, i.e. as the relative contribution of the two flux components vary.

\subsection{Soft intermediate state (SIMS): QPOs without noise} 
\label{sect:hims}

Finally, a last state has been identified, called the Soft Intermediate State (SIMS). This state appears to be particularly important as it has been positively associated the the ejection of the most powerful relativistic jets from this class of systems\cite{fbg2004}. In addition, as I will show in the next sections, it is connected to very fast transitions and shows the highest-frequency features observed to date from black-hole binaries. As its name suggests, this state is somewhere between the HIMS and the HS. It corresponds to a very narrow range of flux characteristics, intermediate between these two states. Indeed, as energy spectral features are concerned, there is no discontinuity between this state and the others. It is in the timing regime that this state stands out: it is characterized by a marked drop in fractional rms, corresponding to the disappearance of the Lorentzian components described above, to be replaced with a very different type of PDS. 

\begin{figure}
   \begin{center}
   \begin{tabular}{c}
   \includegraphics[height=7cm]{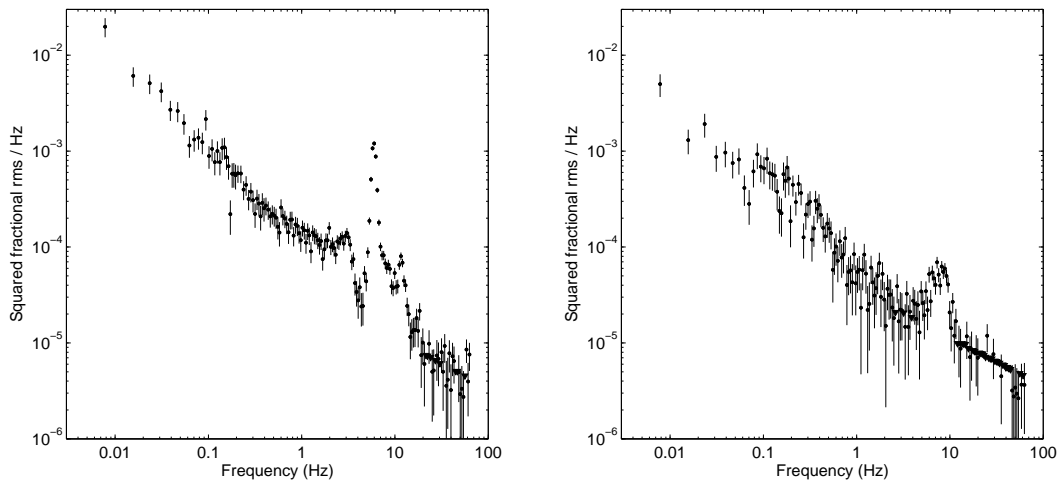}\\
   \end{tabular}
   \end{center}
   \caption[sims_pds] 
   { \label{sims_pds} 
Two PDS from the bright transient XTE J1859--226 when the source was in the Soft-Intermediate state. Left panel: observation from 1999 Oct 18 (integrated 0.01--64 Hz fractional rms 4.6\%). Right panel: observation from 1999 Oct 16 (integrated 0.01--64 Hz fractional rms 2.6\%).
}
   \end{figure}

Two examples can be seen in Fig. \ref{sims_pds}. It is evident that the strong flat-top component of the HIMS is not present, to be replaced by a power law not unlike that observed in the HS (see Fig. \ref{hs_pds}). However, the most striking features in these PDS are the narrow components peaking at a few Hz. Although they are roughly at the same frequency as the strongest QPO peak in Fig. \ref{hims_pds}, their properties are very different, so that they are considered a different class of signal. In one case, the two peaks (that of the HIMS and that of the SIMS) have been seen simultaneously, as a final proof of their different nature. The narrow QPO in the left panel of Fig. \ref{sims_pds} also shows harmonically related peaks at half and twice its frequency, again showing an octave-based harmonic structure. Neither of the two peaks visible in the two panels of Fig. \ref{sims_pds} can be approximated by a Lorentzian shape: a good fit can be obtained with a Gaussian component, with broader external wings for which an additional Lorentzian is necessary. Their frequency is much less variable than that of QPOs in the HIMS: usually it is observed in the 4-8 Hz range.
A more detailed description of these QPOs can be found in the next section.

\section{LOW-FREQUENCY QPO}
\label{sect:lfqpo} 

\begin{figure}
   \begin{center}
   \begin{tabular}{cc}
   \includegraphics[height=7cm]{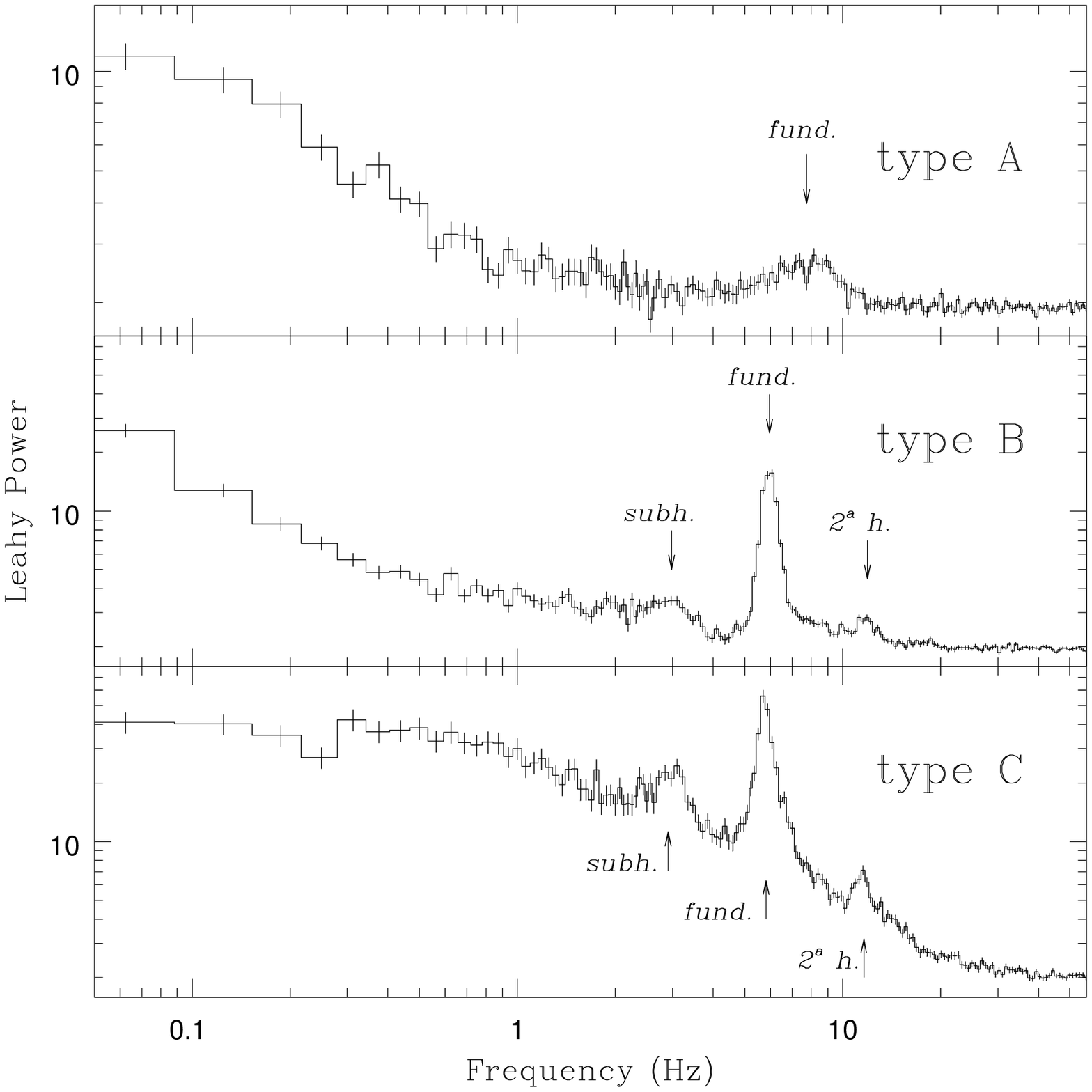} & \includegraphics[height=7cm]{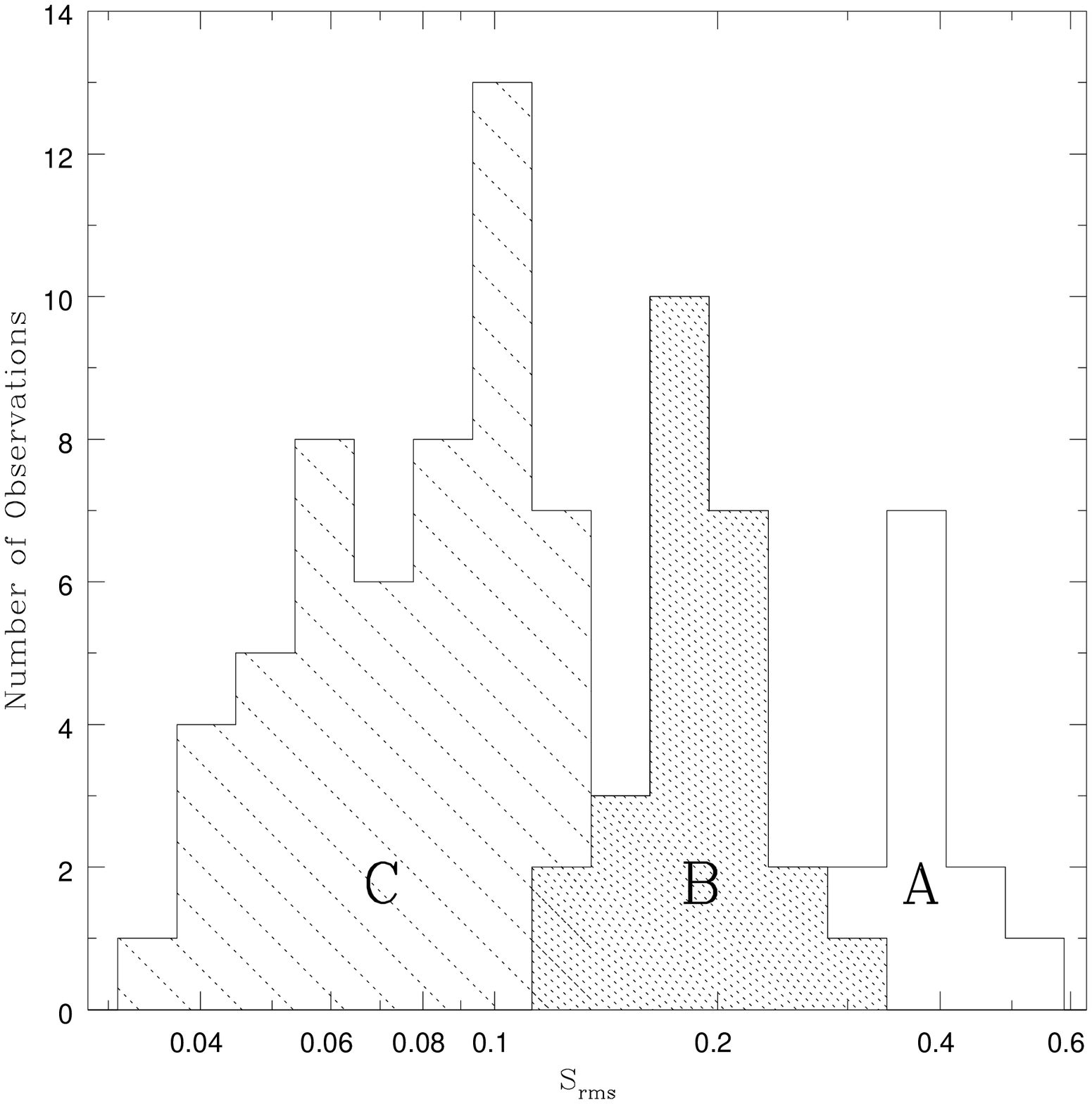}\\
   \end{tabular}
   \end{center}
   \caption[abc_pds] 
   { \label{abc_pds} 
Left panel: examples of the three QPO types\cite{casellaabc}. Here the component due to the Poisson statistics is now subtracted, causing a flattening at high frequencies.
Right panel: Histogram showing the distribution of observed QPO peaks as a function of the inverse of the total fractional rms in the PDS (not only of the QPO)\cite{casellaabc}. The three types of QPO are well separated. The reason why the inverse of the rms is chosen has to do with facilitating comparison with QPOs from neutron-star binaries\cite{casellaabc}.
}
   \end{figure}

In the previous section, I have shown examples of PDS of the different states observed in black-hole binaries. While broad-band noise, when present, accounts for most of the observed aperiodic variability, the most striking features are the QPOs observed in some of the states. The reason is clear: a more coherent peak provides a characteristic frequency which can be associated with a physical quantity in the accretion flow. All QPO peaks shown above have frequencies in the range 0.1--15 Hz and are commonly referred to as ``low-frequency QPO''. With a few exceptions, these QPOs are only seen in the two Intermediate states, which are the shortest living ones. In addition to their astrophysical interest, these features are the most interesting signals to study. 
In the recent years, it has become clear that these QPOs are not a homogeneous set and can be classified into three very separate types with very different properties. These types are (rather synthetically) called A, B and C\cite{wijnands1999,remillard2002,casellaabc}. Three examples can be seen in the left panel of Fig. \ref{abc_pds}. In addition to the detailed characteristics described below, the three types can be separated by simply considering the {\it total} fractional rms variability of their PDS, including the noise components\cite{casellaabc}. In the right panel of Fig. \ref{abc_pds} is shown the distribution of QPO detections as a function of the inverse of the total fractional rms. 

\subsection{Type-C QPO} 
\label{sect:typec}

Type-C QPOs correspond to the peaks shown in Fig. \ref{hims_pds}, observed when the source is in the HIMS. At very-low frequency, they are also sometimes seen in the LS (see the small peak in Fig. \ref{ls_pds}). They are strong signals (up to 16\% rms), they are rather narrow (Q$\sim$7--12) and they are quite variable, both in frequency and intensity. Their centroid is in the range 0.1--15 Hz and it is superposed on a flat-top noise component that steepens at a frequency comparable to that of the QPO (se Fig \ref{hims_pds}). Overall, the noise plus QPO variability can be as high as 30--40\%. 
As described above, their frequency correlates positively with the source flux.
A sub-harmonic and a second harmonic peak (plus rarely a fourth harmonic peak) can be usually seen (see Fig. \ref{abc_pds}). The integrated fractional rms of the QPO increases with energy, flattening above 10 keV. This QPO is rather ubiquitous, being observed in almost all black-hole binaries.

\subsection{Type-B and type-A QPO} 
\label{sect:typeb}

A type-B QPO can be seen in the left panel of Fig. \ref{sims_pds}. They correspond to peaks with a typical fractional rms of $\sim$4\% and a Q$\sim$6. They are detected {\it exclusively} in the SIMS. Unlike the type-C peaks, they are found in a narrow range of frequencies, between 4 and 8 Hz. Also, they are characterized by the absence of a strong flat-top noise component, but are associated to a weaker power-law continuum component. As it is evident from the left panels of Fig. \ref{sims_pds} and Fig. \ref{abc_pds}, a sub-harmonic and a second harmonic are also usually present, once again presenting the octave-based harmonic structure. Sometimes, unlike for type-C QPOs, the subharmonic is as strong as the fundamental\cite{casella1859}.

Type-A QPOs are the most elusive, as they are not much coherent and rather weak. An example can be seen in the right panel of Fig. \ref{sims_pds}. As for type-B QPOs, they are only observed in the SIMS. The integrated fractional rms is a few \% and the Q parameter is less than 3. Their centroid frequency is always around 8 Hz. Being seen in the SIMS, they are associated to a weak continuum component in the PDS. No (sub-)harmonic  have been detected. Sometimes, type-A QPOs are so weak that they can be detected only by adding up the data from many observations.

\subsection{Fast transitions} 
\label{sect:fast_transitions}

\begin{figure}
   \begin{center}
   \begin{tabular}{cc}
   \includegraphics[height=6cm]{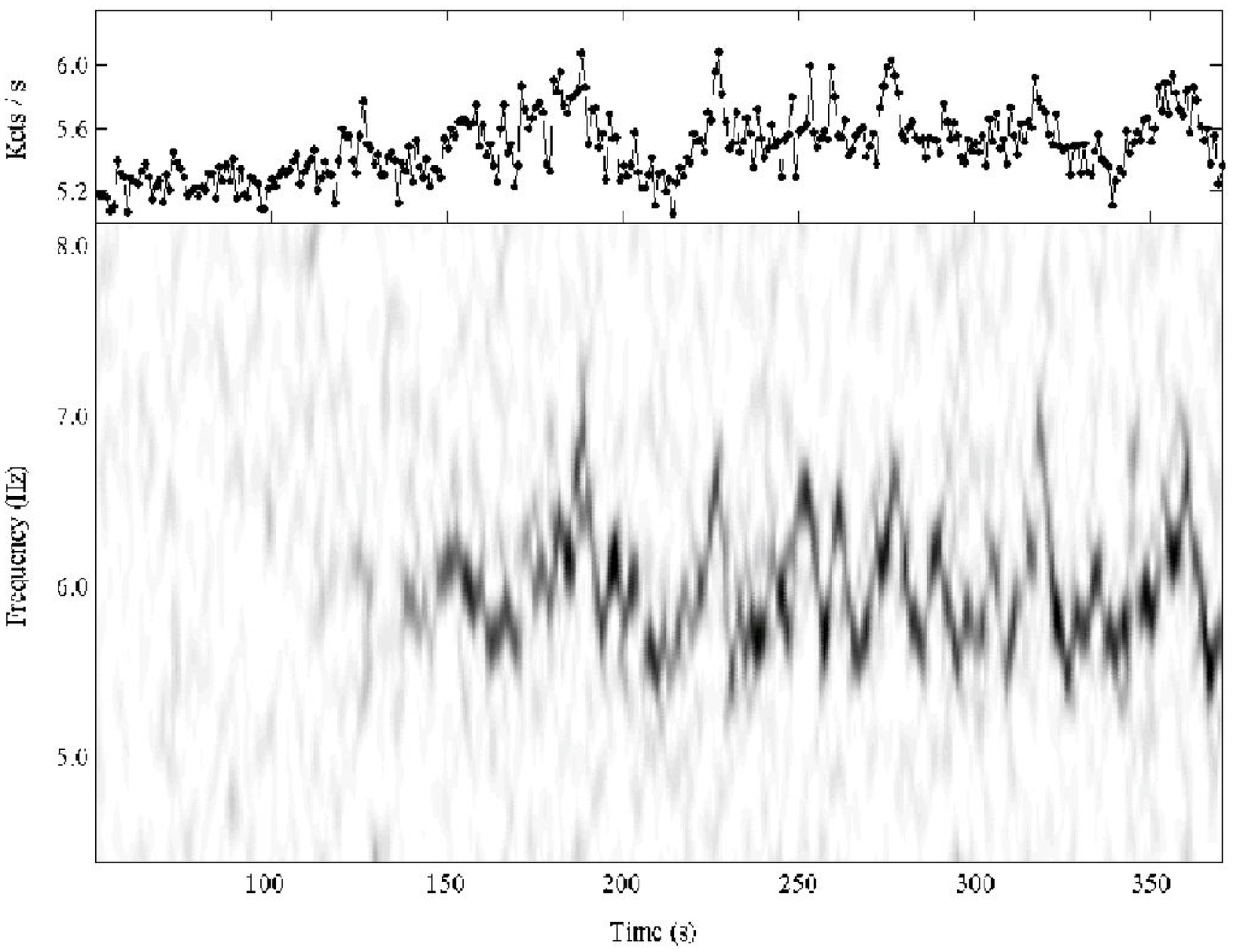} & \includegraphics[height=6cm]{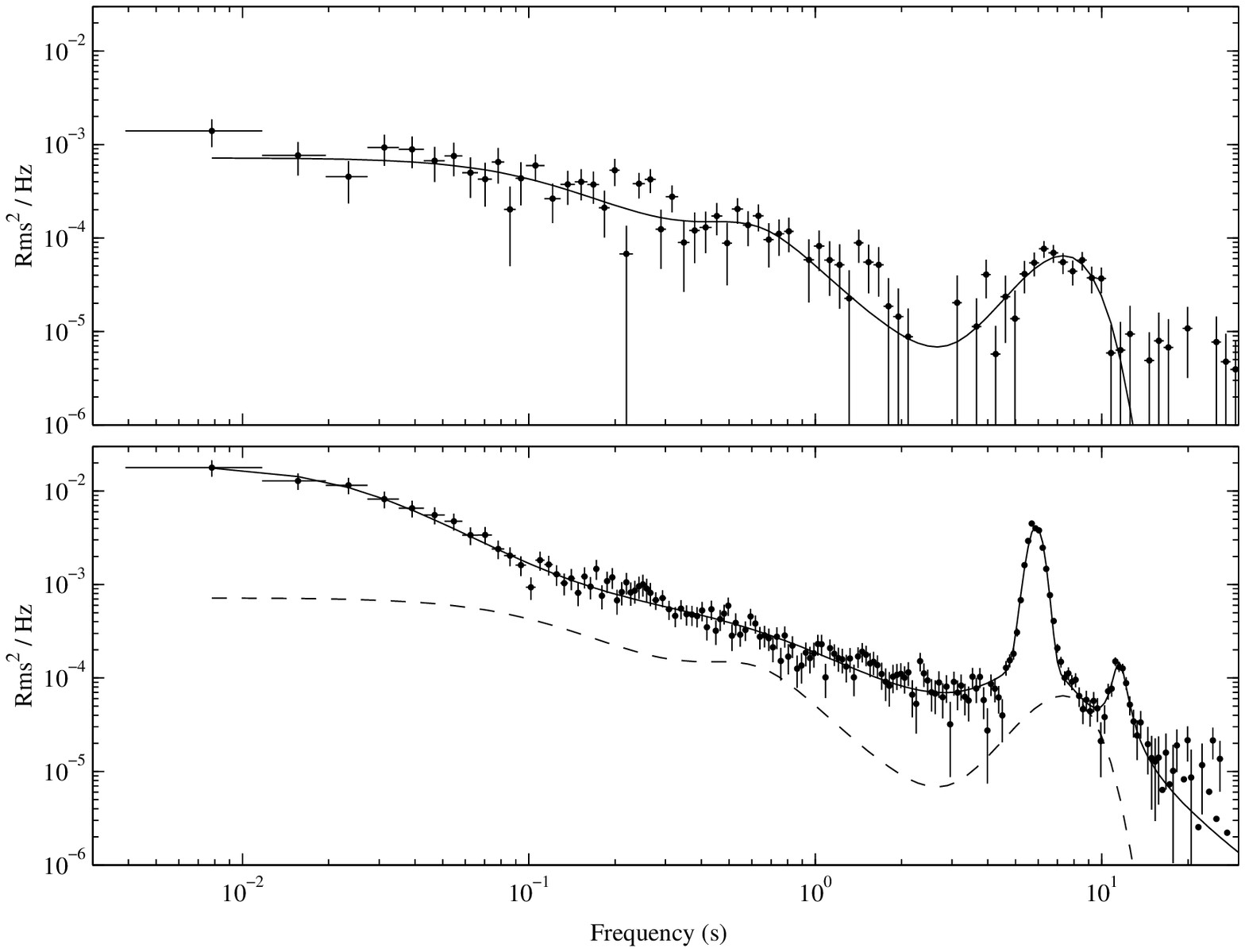}\\
   \end{tabular}
   \end{center}
   \caption[trans_pds] 
   { \label{trans_pds} 
Left panel: spectrogram of a segment of a SIMS observation of GX 339--4, zoomed on the 4.5--8 Hz frequency region. On the top, the corresponding flux evolution of the source is shown, one point per column of the spectrogram\cite{nespoli339}. 
Right panel: integrated PDS of the first 120 seconds (top) and of the rest of the spectrogram shown in the left panel (bottom). The dashed line duplicates the model PDS of the first 120 seconds\cite{nespoli339}.
}
   \end{figure}

The low-frequency QPOs described above are very transient features. An example of this can be seen in the left panel of Fig. \ref{trans_pds}, which shows a spectrogram of about six minutes of observation of GX 339--4 when in the SIMS\cite{nespoli339}. There is an obvious narrow feature at 6 Hz, which can be identified with a type-B QPO (see bottom-right panel in Fig. \ref{trans_pds}). Two things are noticeable: the first is that the centroid frequency of the QPO is quite variable on a short time scale. The variations are not random: a Fourier analysis of the frequency evolution shows that it changes on a typical time scale of 10 seconds. The second is that the QPO is not visible in the first $\sim$2 minutes of data. The average PDS of those 120 seconds shows that a type-A QPO is present, not visible in the spectrogram due to its weakness (see top-right panel in Fig. \ref{trans_pds}). 
Many of these fast transitions, have been observed in different systems 
\cite{miyamoto339,takizawa,bel3392005,casella1859}, 
involving rapid changes from one QPO type to another. Clearly, the fast-variability properties of the accretion flow can change on a very short time scale. Interestingly, the corresponding changes in the flux spectral distribution are minor, often undetectable. For the case shown in Fig. \ref{trans_pds}, from the left panel an increase in the source flux can be seen in correspondence to the transition, but the energy spectrum does not seem to vary within errors. More recently, there has been evidence of the fact that there are stronger variations associated to the transition, but only in photons at higher energies, not detectable by the RXTE/PCA instrument\cite{bel3392006}.

While type-A QPOs are difficult to study due to their weakness, it is interesting to compare in detail the properties of the other two types. While the type-C QPO is observed over a wide range of frequencies, roughly from 0.1 to 15 Hz and correlated with the source flux distribution, its frequency is stable on a short time scale and does not show anything like the wild variations visible in the left panel of Fig. \ref{trans_pds}. The type-B QPO varies over a much reduced range of frequencies, 4--8 Hz but is always observed to jitter in time on short time scales.
Although it is often ignored, these two oscillations are extremely different observationally and most likely related to very different physical mechanisms.

\section{HIGH-FREQUENCY QPO}
\label{sect:hfqpo}

In addition to low-frequency QPO, additional peaks at higher frequency (30--450 Hz) have been observed in a few cases (see Ref. \cite{ronjeff} but also ref. \cite{bellonihf1915}).
. Their properties are markedly different from their low-frequency counterpart. They are very elusive signals, which have been observed in few observations of a few systems.
Currently, single high-frequency peaks have been detected in three systems, while in four other sources {\it two} high-frequency QPOs are known.  As mentioned, these features are detected only a few times. They are intrinsically weak, with an integrated fractional rms of 1--3\%, which however is strongly dependent on the energy band considered for the analysis. 
In most systems, the centroid frequency of the high-frequency QPO appears rather stable, showing only small variations between observations. This is in marked contrast with the low-frequency QPOs. 
The fact that this QPO is observed more or less at the same frequency in a given source suggests that its centroid is driven by fundamental parameters of the system. Since their frequency is in the correct range to reflect the orbital keplerian frequency in the innermost parts of the accretion disk, is has been speculated that from it we could, once known the mass of the central black hole through other methods, estimate the value of the angular momentum of the compact object, the second and most elusive physical parameter of astrophysical interest. However, a full theoretical model for these QPOs is not available yet. 

\begin{figure}
   \begin{center}
   \begin{tabular}{cc}
   \includegraphics[height=7cm]{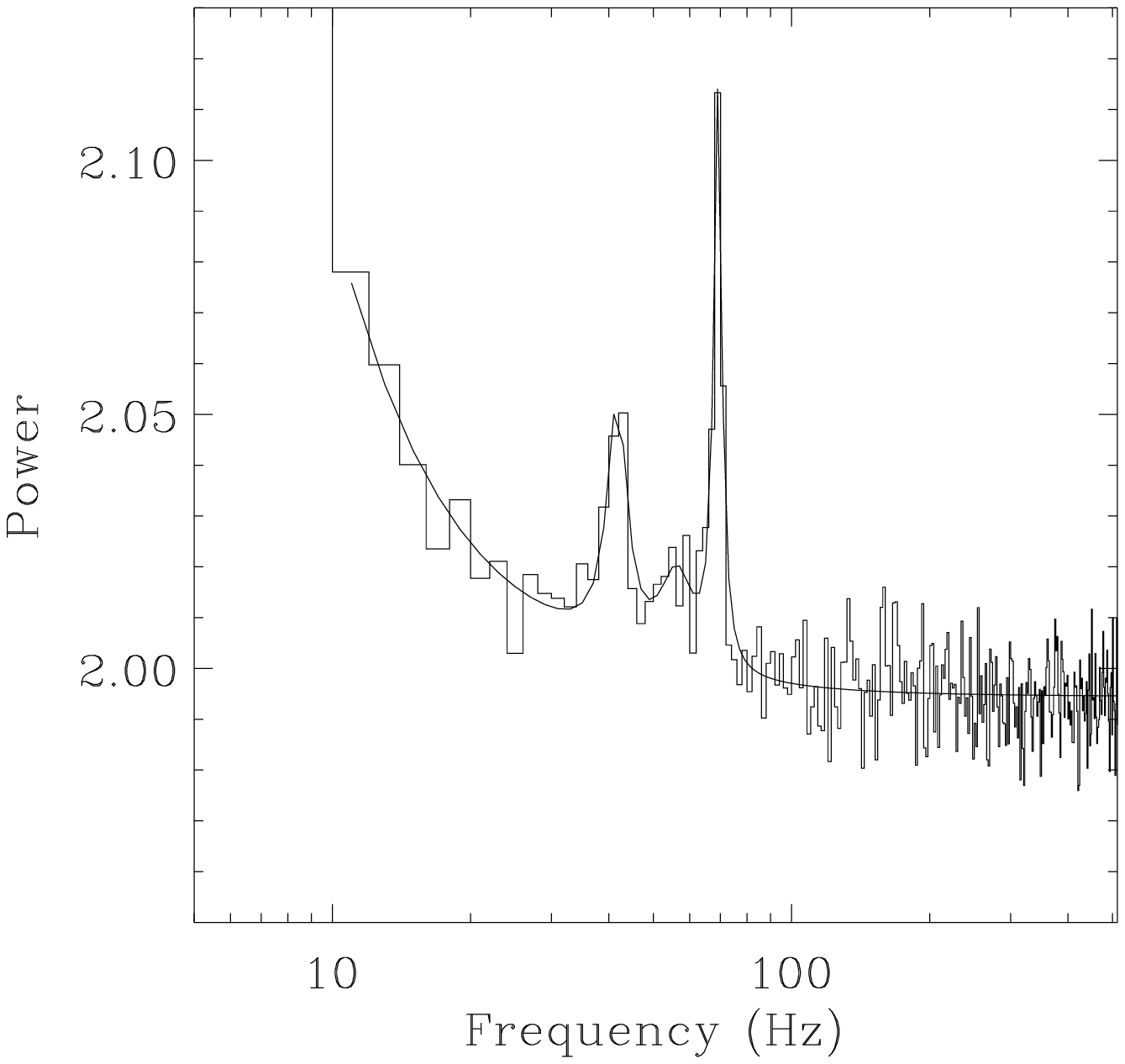} & \includegraphics[height=7cm]{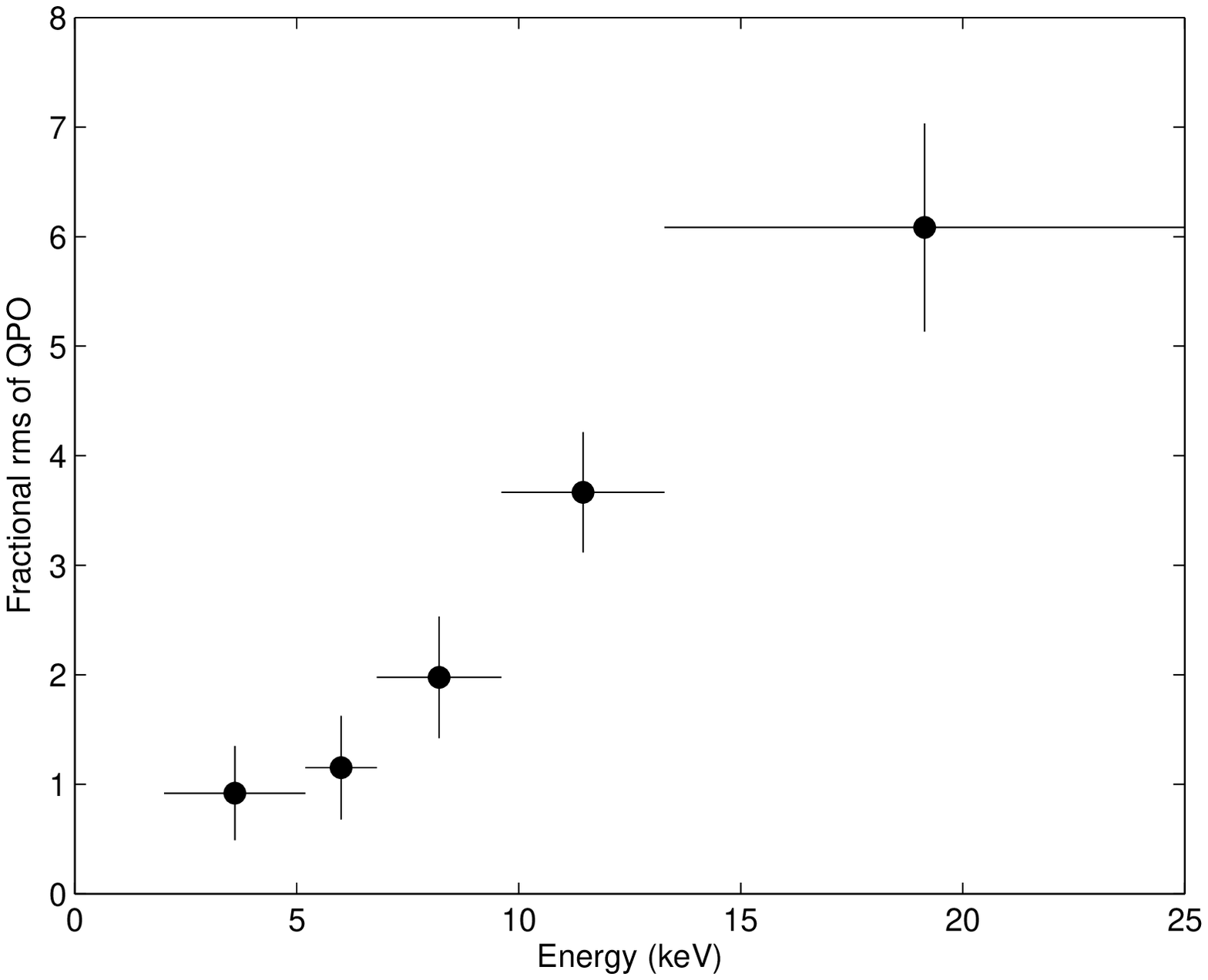}\\
   \end{tabular}
   \end{center}
   \caption[hfqpo_pds] 
   { \label{hfqpo_pds} 
Left panel: twin high-frequency QPO in a RXTE observation of GRS 1915+105\cite{tod1915}. The component due to the Poissonian noise was not subtracted from the PDS.
Right panel: energy dependence of the integrated fractional rms of the 69 Hz peak in GRS 1915+105, measured in the only observation when the feature was particularly strong\cite{morgan1915}.
}
   \end{figure}

It has been noted\cite{abrklu2001} that when two QPO peaks are observed not only their frequencies are stable but they are consistent with a simple ratio (2:3 or 3:5). An example of such a pair of QPOs can be seen in the left panel of Fig.\ref{hfqpo_pds}, where clear peaks at 41 Hz and 69 Hz are seen
\cite{tod1915} in the PDS of an RXTE observation of GRS 1915+105. Interestingly, there are other RXTE observations of the same source taken a few days before or after these detections, where all other observable parameters are the same, but the QPOs are not detected. The right panel of Fig.
\ref{hfqpo_pds} shows the integrated fractional rms of the 69 Hz peak in GRS 1915+105 as a function of energy: the signal is as weak as 1\% rms below 5 keV, but increases its strength to more than 6\% above 15 keV. Unfortunately, this is the strongest detection we have, so that such a plot for other sources cannot be obtained.
With only one exception, all high-frequency QPO peaks have been detected when the source was in the SIMS\cite{bellonihf1915}. This is important, as it makes them mutually exclusive with type-C QPOs (the only, unpublished, case of HIMS detection of a high-frequency QPO was in an extreme HIMS, very close to the SIMS). Given all correlations between frequencies of PDS components in the LS and the HIMS, this means that high-frequency QPOs, whose centroid frequency is almost constant, are associated to type-A/B QPOs, whose centroid frequency is not much variable.

Although we know very few cases, these preferred ratios appear important for our understanding of the high-frequency signals and for its connection to parameters of fundamental physics. For instance, it has been suggested that the two observed frequency are due to a resonance between the orbital and the radial epicyclic motion of matter in near-Keplerian orbit around the black hole\cite{abrklu2001}.
In General Relativity, a particle in orbit around a mass on an elliptical orbit will display a precession of its orbit, which is responsible for instance for the precession of the perihelion of Mercury. The model predicts that when the orbital and the epicyclic frequency resonate, observable signals will be detected. This would be important, as for black-holes of measured mass this would lead to the determination of its spin, the second elusive parameter of such an object.

\section{CONCLUSIONS}
\label{sect:conclusions} 

The panorama I presented in the previous section is very complex. It is clear that fast aperiodic variability, in the form of broad-band noise and QPOs, is an important aspect of the emission from accretion of matter onto a black hole, but at the time of writing we do not have a good physical interpretation for what we see. The frequencies and the energies involved indicate that we are dealing with processes in the very inner parts of the accretion flow, very close to the black hole or, better, to the innermost stable orbit around it, which is predicted by General Relativity. There is a possible interpretation for the high-frequency QPOs, which are the features about which we have the least information, but no successful interpretation for all other components and frequencies exist to date.

   \begin{figure}
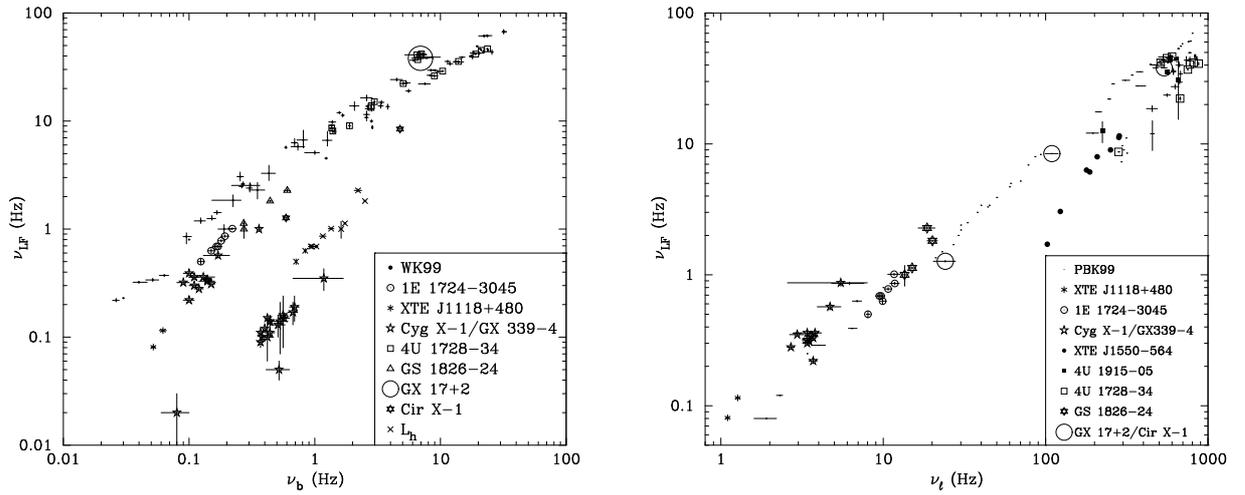

   \begin{center}
   \begin{tabular}{cc}
   \includegraphics[height=7cm]{wk.ps} & \includegraphics[height=7cm]{pbk_low.ps}\\
   \end{tabular}
   \end{center}
   \caption[corr_pds] 
   { \label{corr_pds} 
Left panel: correlation between the type-C QPO (and its analogous for neutron-star binaries) and the characteristic frequency of the flat-top noise component\cite{wk,bpk}. Neutron-star binaries occupy  the upper right of the correlation.
Right panel: correlation between the type-C QPO (and its analogous for neutron-star binaries) and selected features at higher frequencies\cite{pbk,bpk}. For details see \citenum{pbk}.
}
   \end{figure} 

One important aspect which I did not touch is the connection to the variability observed from neutron-star binaries, very similar astronomical systems where the compact object is a neutron star. As mentioned above, neutron stars have a solid surface and usually a non-neglgible magnetic field, both properties that are likely to influence the observational properties. Indeed, as can be seen in the article by Michiel van der Klis in this volume, there are strong differences. However, there are also major similarities. 
In 1999 it was found that the characteristic frequencies of the low-frequency components of black-hole binaries and other frequencies in neutron-star binaries follow very tight correlations between them, regardless of the nature of the compact object\cite{wk,pbk,bpk,vdk2006}. 
Two such correlations can be seen in Fig. \ref{corr_pds}. On the left panel is a correlation between the low-frequency type-C QPO (Y axis) and the lowest frequency, corresponding to the turnover to flat-top at low frequencies (see Figs. \ref{ls_pds} and \ref{hims_pds})\cite{wk}. Black holes occupy the leftmost part of the correlation, neutron stars the upper one. This is consistent with the different masses of the objects. Black holes have a typical mass seven times that of neutron star, corresponding to slower dynamical time scales by the same amount.
On the right panel is a correlation between the same low-frequency type-C QPO (Y axis) and higher-frequency features, which can be broad Lorentzians for black holes or narrow kHz QPOs for neutron stars (see van der Klis, this volume).
These correlations indicate that, although with differences, the process underlying these variability process is most likely the same, independent of the compact object in the center of the accretion flow. 

The correlation in the right panel of Fig. \ref{corr_pds} has been interpreted in terms of the so-called ``Relativistic Precession Model''\cite{stellavietri,stellavietrimorsink}, in which the main observed frequencies are related to fundamental frequencies derived from General Relativity. Although this model does not explain all observed features, in particular in the case of neutron-star binaries, it focuses attention on relativistic frequencies and has opened the way to other models, such as that for high-frequency QPOs mentiond above. In particular, there has been an attempt to connect these frequencies to a more physical process of production of actual signals\cite{psaltisnorman} at the observed frequencies. 

The RXTE satellite will be operative only until 2008 March, after which there will be no other instrument capable of accumulating time series of the required quality for these studies. Typical modern instruments for X-ray astronomy are based on very sensitive detectors with low-background. This means not only that the typical number of photons per second accumulated from the objects of our interest are not sufficient for our analysis, but also that they are too many for the detector itself. This means that it will be very difficult even to observe, not to mention to reach the required sensitivity. 
Plans for an astronomical mission if not dedicated, at least partially optimized for high-sensitivity timing studies are necessary. For the time being, the enormous database provided by RXTE is available for analysis and interpretation, so that new results from it will keep coming even after the end of the mission.

\acknowledgments     
 
The author acknowledges PRIN INAF 2006 grant.


\bibliography{report}   
\bibliographystyle{spiebib}   


\end{document}